\documentclass[twocolumn,twocolappendix]{aastex631}

\usepackage{apjfonts}
\usepackage{amsmath}
\usepackage{amssymb}
\usepackage{bm}
\usepackage{booktabs}
\usepackage{graphicx}
\usepackage{color}

\newcommand{\vect}[1]{\mathbf{#1}}
\newcommand{\uvect}[1]{\hat{\mathbf{#1}}}

\shorttitle{Dual-Burst Rings}
\shortauthors{Seto}

\journalinfo{}

\begin{document}
\renewcommand{\theequation}{\thesection.\arabic{equation}}
\renewcommand{\theHequation}{\thesection.\arabic{equation}}
\title{A Dual-Burst Geometrical Prescription for Concurrent Signaling}

\author{Naoki Seto}
\affil{Department of Physics, Kyoto University, Kyoto 606-8502, Japan}
\email{seto@tap.scphys.kyoto-u.ac.jp}

\begin{abstract}
We propose a dual-burst implementation of concurrent signaling for
technosignature searches.  Concurrent signaling is a Schelling-point
prescription for implicit coordination between transmitters and
receivers without prior communication, using salient astronomical
phenomena as coordination anchors.  It allows possible transmitters at
different line-of-sight distances to be searched collectively through a
time-dependent locus on the sky.  In the dual-burst implementation, the
coordination anchors are two cosmological transients.  At a chosen
observing time after the later burst, the leading geometrical
prescription specifies a precisely defined sky ring from the two burst
directions and their observed arrival times at the receiver.  No
distance to either burst, to a candidate transmitter, or to a Galactic
spatial anchor is required to construct this angular search locus.  The
finite width of the ring is therefore controlled primarily by burst
localization rather than by an astrophysical distance scale, thereby
reducing the set of sky directions to be searched.

\end{abstract}


\section{Introduction}
\label{sec:intro}

A central difficulty in searches for intentional interstellar signals is
the size of the signaling parameter space.  A receiver must choose not
only where to look on the sky, but also when to look, over what
frequency or wavelength range, and for what class of signal morphology
\citep[e.g.,][]{Tarter2001,Horowitz1993,Wright2020}.  The transmitter
faces a related problem: without prior coordination, it is inefficient
to transmit arbitrarily over the full space of directions, times, and
signal types.  An efficient strategy should therefore exploit
references and rules that both sides can identify independently.

This is the role of a Schelling point.  In the present context, a
Schelling point is not merely a conspicuous object, but a basis for
implicit coordination in a shared strategy space \citep{Schelling1960}.
Useful prescriptions should therefore be simple and reproducible,
preferably tied to salient astronomical events or symmetric geometrical
structures rather than to arbitrary private choices.  Such a prescription
can guide both when transmitters send and when receivers search.

Concurrent signaling is one way to implement this idea
\citep{Seto2019,Seto2021,Seto2024}, and is closely related to
earlier timing-based SETI strategies that use conspicuous astronomical
events.  In early SETI Ellipsoid work, a single transient event and the
distance to a target star define a predicted reception epoch for a
signal timed to that event, making accurate target distances an
essential part of the prescription
\citep{1977Icar...32..464M,1980Icar...41..178M,1994Ap&SS.214..209L}
(see also \citealt{Davenport2022,2023AJ....166...79N,Cabrales2024} for
recent observational studies).  By contrast, in a concurrent-signaling scheme, a given sky direction is
searched at a specified observing epoch.  The search then covers possible
transmitters in that direction over a range of line-of-sight distances,
rather than being tied to a single target distance.  This is achieved by assigning, for each possible
signal trajectory, a common reference point and a common reference time
that can be determined independently by transmitters and receivers.  The
transmitter-side and receiver-side tasks are then two sides of the same
prescription: a transmitter decides whether and when to emit relative to
its assigned reference event, while the receiver applies the same rule to
decide where and when to search.  The key problem is therefore how to
define these common reference points and times using astronomical
information available to both sides.

Earlier implementations realized this assignment by using a
finite-distance astronomical reference.  In schemes based on a single
finite-distance anchor, the reference event itself had to provide both a
well-defined epoch and a three-dimensional position.  A later hybrid
strategy reduced this burden by using an extragalactic burst as a
temporal marker and the Galactic center as a spatial reference
\citep{Seto:2025iul}.  However, any prescription tied to a
finite-distance spatial anchor retains a distance-error term and a
corresponding reference-selection ambiguity.  A prescription based only
on timing and direction would have a cleaner error hierarchy and would be
a natural target for implicit coordination among independently acting
participants.

\begin{figure*}
    \centering
    \includegraphics[width=1.5\columnwidth]{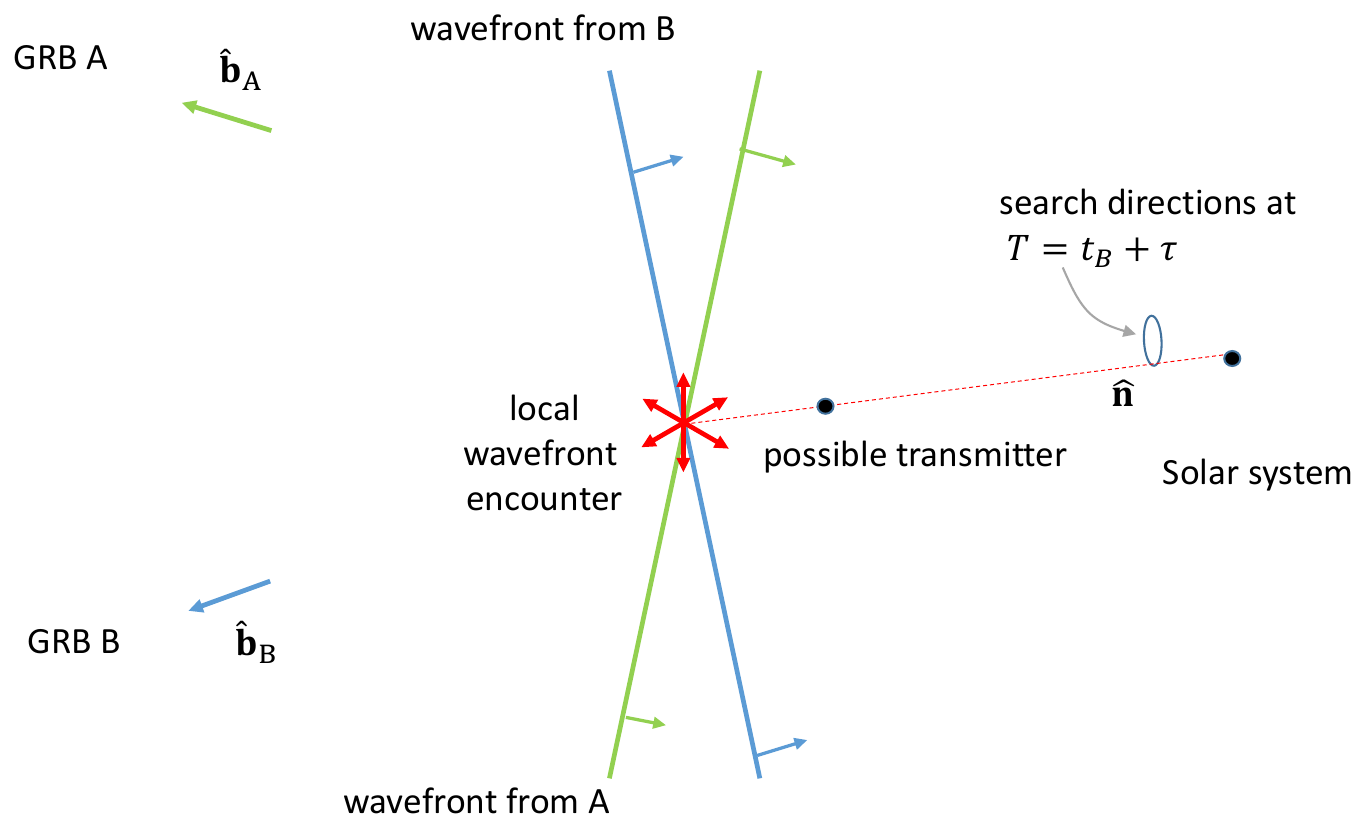}
   \caption{
Schematic illustration of the dual-burst virtual-emission prescription.
Two distant bursts, A and B, are observed by the receiver, here taken to
be the Solar System, at times \(t_A\) and \(t_B\), and are represented by
approximately planar wavefronts associated with the unit source
directions \(\uvect{b}_A\) and \(\uvect{b}_B\).  Each local encounter of
the two wavefronts defines a reference event of the prescription.  The
encounter shown here is only a representative example; in
three-dimensional space, the set of such encounters forms a
two-dimensional surface.  Isotropic outgoing virtual rays are assigned
to these reference events.  A possible transmitter located on one such ray may emit an artificial
signal along that virtual ray.
For an observing time \(T=t_B+\tau\), a receiver applies the same
prescription to determine the corresponding arrival directions
\(\uvect{n}\) on the sky, forming the search ring derived in
Section~\ref{sec:prescription}. 
In the plane-wave approximation, this construction depends only on the
two burst directions and their observed arrival times, not on the
distances to the bursts, to the transmitter, or to any Galactic spatial
anchor.
}
    \label{fig:concept}
\end{figure*}

The present paper removes the finite-distance spatial reference by using
two distant bursts.  A single distant burst provides a salient time
marker and sky direction for each observer, but it does not by itself
define spacetime reference events for signal trajectories across the
Galaxy.  Two distant bursts provide the minimal additional structure:
the local encounters of their wavefronts define such reference events
across the Galaxy, as illustrated in Figure~\ref{fig:concept}.  We use
these encounters as virtual-emission events, to which outgoing virtual
rays with their associated emission times are assigned.  A transmitter
may emit an artificial signal along one of these assigned virtual rays,
while a receiver applies the same prescription to determine the
corresponding arrival directions at a chosen observing time. The construction is geometrical rather than tied
to a particular coordinate frame; a receiver frame is introduced below
only to compute the resulting search ring.

In the plane-wave approximation, the receiver-side search directions,
and hence the predicted sky ring, depend only on the two burst
directions and their observed arrival times.  No distance to either
burst, to a candidate transmitter, or to a Galactic spatial anchor is
required.  The predicted search ring is therefore controlled primarily
by the angular localizations of the two bursts rather than by an
astrophysical distance scale.  Although we use gamma-ray bursts (GRBs)
as the main working example, the same construction applies to any pair
of sufficiently distant, conspicuous, and well-localized transients with
well-defined arrival times \citep[see also][]{Corbet:1999vv}.

In this paper we develop the geometry of this dual-burst prescription.
We first formulate the virtual-emission rule and derive the exact
plane-wave search ring for an arbitrary observing time after the later
burst.  We then examine the ring geometry, its small-angle form,
and the associated virtual-emission depth scale.  After estimating the
leading error budget, we discuss burst-pair selection as the remaining
Schelling-like problem.  We do not perform a survey-data analysis here;
the goal is to provide a compact geometrical prescription that can be
applied to archival or future time-domain SETI searches.

\section{Dual-burst prescription and search ring}
\label{sec:prescription}

We now turn the dual-burst idea into an explicit search prescription for
a receiver at the Solar System.  As illustrated in
Figure~\ref{fig:concept}, the local encounters of the two burst
wavefronts define virtual reference events across the Galaxy.  
To each event we assign outgoing virtual rays, understood as lightlike
trajectories with their corresponding virtual-emission times.  A
transmitter located on one such ray may emit an artificial signal along
the assigned virtual ray, while a receiver applies the same prescription
to determine which arrival directions should be searched at a chosen
observing time.  We now derive the corresponding search ring.

Consider two distant bursts, A and B, observed at the Solar System at
times \(t_A\) and \(t_B\), respectively, with \(t_B>t_A\).  Let
\(\Delta\equiv t_B-t_A>0\), and let \(\uvect{b}_A\) and
\(\uvect{b}_B\) be the corresponding source directions on the sky,
pointing from the Solar System toward the two bursts.  We denote their
angular separation by
\(\cos\alpha=\uvect{b}_A\cdot\uvect{b}_B\).  For an observing time
\(T=t_B+\tau\), define the dimensionless delay \(u=\tau/\Delta\).

We work in a Solar-System-barycentric inertial frame, with the Solar
System at \(\vect{x}=0\) and coordinate time \(t\).  In the plane-wave
approximation, the wavefronts of bursts A and B are represented by
\begin{equation}
    t-t_A+{\uvect{b}_A\cdot\vect{x}\over c}=0,
    \qquad
    t-t_B+{\uvect{b}_B\cdot\vect{x}\over c}=0 .
    \label{eq:wavefronts}
\end{equation}

Now consider a virtual ray whose corresponding signal is received at
the Solar System at
\begin{equation}
    T=t_B+\tau,\qquad \tau>0 .
\end{equation}
If it arrives from sky direction \(\uvect{n}\), with \(\uvect{n}\) a unit
vector pointing from the Solar System along the arrival direction, and
the associated virtual-emission event is at distance \(s\) along that
direction, then the event is
\begin{equation}
    (t,\vect{x})
    =
    \left(
    t_B+\tau-{s\over c},\,
    s\uvect{n}
    \right).
    \label{eq:emission_event}
\end{equation}
The dual-burst prescription requires this event to lie on both burst
wavefronts.  Substituting Eq.~\eqref{eq:emission_event} into
Eq.~\eqref{eq:wavefronts} gives
\begin{equation}
    \Delta+\tau
    =
    {s\over c}
    \left(1-\uvect{b}_A\cdot\uvect{n}\right),
    \qquad
    \tau
    =
    {s\over c}
    \left(1-\uvect{b}_B\cdot\uvect{n}\right).
    \label{eq:two_conditions}
\end{equation}
Eliminating \(s\), we obtain the plane-wave search-ring equation
\begin{equation}
    \left[
    (1+u)\uvect{b}_B-u\uvect{b}_A
    \right]\cdot\uvect{n}=1 .
    \label{eq:ring_exact}
\end{equation}
At a fixed observing time \(T=t_B+\tau\), this equation restricts the
unit arrival direction \(\uvect{n}\) to the intersection of the unit
sphere with a plane.  As shown explicitly in the next section, this
locus is a small circle, or search ring.

Equation~\eqref{eq:ring_exact} is the central plane-wave result of the
construction.  The predicted search ring depends only on the burst
directions \(\uvect{b}_A\), \(\uvect{b}_B\), the observed arrival-time
separation \(\Delta\), and the elapsed observing time \(\tau\).  It does
not require the distances to the bursts, the distance to a possible
transmitter, or the distance to any Galactic spatial anchor.  This
distance independence follows from treating the distant-burst wavefronts
as plane waves; for cosmological bursts, we show below that the
corresponding curvature correction is smaller than the angular
localization scale relevant for the searches considered here.

The same equations also give the line-of-sight distance to the
corresponding virtual-emission event.  From the second of
Eqs.~\eqref{eq:two_conditions},
\begin{equation}
    s(\uvect{n},\tau)
    =
    {c\tau\over 1-\uvect{b}_B\cdot\uvect{n}} .
    \label{eq:s_of_n}
\end{equation}
Thus the search direction \(\uvect{n}\) and the associated
virtual-emission depth \(s\) are both determined by timing and angular
information alone.

Equation~\eqref{eq:s_of_n} is written for the Solar System endpoint,
but it reflects a more general causal ordering along the assigned
virtual ray.  At any point on that ray, the time interval between the
local arrival of the later burst B and the arrival of the 
virtual photon is proportional to the path length from the corresponding
local wavefront encounter to that point.  This interval is therefore
non-negative, and vanishes only at the encounter itself.  A transmitter
on the ray can therefore emit after receiving both burst wavefronts at
its own location, rather than anticipating the later burst.

\section{Search-ring geometry and depth scale}
\label{sec:ring_depth}

We now examine the geometry of Eq.~\eqref{eq:ring_exact} for \(u\ge0\).
It is useful to define
\begin{equation}
    \vect{q}(u)
    \equiv
    (1+u)\uvect{b}_B-u\uvect{b}_A .
    \label{eq:q_def}
\end{equation}
Then the exact search condition is
\begin{equation}
    \vect{q}(u)\cdot\uvect{n}=1 .
    \label{eq:q_ring}
\end{equation}
This is the intersection of the unit sphere with a plane, and therefore
defines a small circle on the sky.  Its center direction is the direction
of \(\vect{q}(u)\), and its angular radius \(\rho\) is determined by
\begin{equation}
    \cos\rho(u)
    =
    {1\over |\vect{q}(u)|}.
    \label{eq:ring_radius_exact}
\end{equation}
Using \(\cos\alpha=\uvect{b}_A\cdot\uvect{b}_B\), we obtain
\begin{equation}
    |\vect{q}(u)|^2
    =
    1+2u(1+u)(1-\cos\alpha) \ge 1.
    \label{eq:q_norm}
\end{equation}
At \(u=0\), Eq.~\eqref{eq:q_def} gives
\(\vect{q}=\uvect{b}_B\), so the ring is collapsed in the direction of
burst B.  As \(u\) increases, the term \(-u\uvect{b}_A\) shifts the
center direction \(\vect{q}(u)\) from B toward the side opposite to A on
the sky.  For any nonzero burst separation, Eq.~\eqref{eq:q_norm} shows
that \(|\vect{q}(u)|\) increases monotonically with \(u\).  Since
\(\cos\rho=1/|\vect{q}|\), the angular radius also grows monotonically
with observing time.

\begin{figure}
    \centering
    \includegraphics[width=0.5\textwidth]{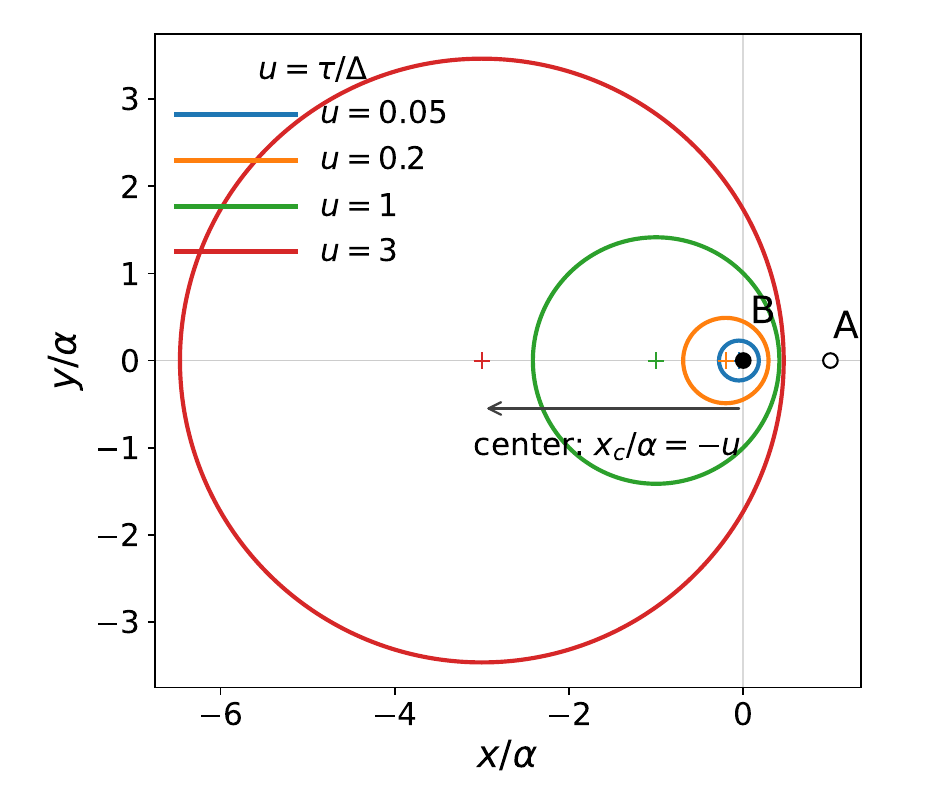}
    \caption{
Small-angle illustration of the search-ring evolution in the tangent
plane centered on burst B.  Coordinates are measured in units of the
burst separation \(\alpha\), with B at the origin and A at
\(x/\alpha=1\).  At dimensionless observing delay \(u=\tau/\Delta\), the
ring center is displaced to \(x/\alpha=-u\), while the ring radius is
\(\rho_{\rm ring}/\alpha=\sqrt{u(1+u)}\).  Thus the ring center moves
from B toward the side opposite to A, while the ring radius grows with
time.
}
\label{fig:ring-evolution}
\end{figure}

In the small-angle limit \(\alpha\ll1\), this motion has a particularly
simple form.  Project the sky near burst B onto a tangent plane, and
place burst A at angular displacement \(\vect{a}\) from B, with
\(|\vect{a}|=\alpha\).  Then the ring center is displaced from B by
\(-u\vect{a}\), while the ring radius is
\begin{equation}
    \rho_{\rm ring}
    =
    \alpha\sqrt{u(1+u)} .
    \label{eq:small_angle_radius}
\end{equation}
Thus, in units of the burst separation, the center displacement is
\(u\), and the radius is \(\sqrt{u(1+u)}\).  This is the geometry
illustrated in Figure~\ref{fig:ring-evolution}.

The associated line-of-sight depth is given by Eq.~\eqref{eq:s_of_n}.
At a fixed observing time, the sky directions \(\uvect{n}\) are
restricted to the search ring, but the quantity
\(\uvect{b}_B\cdot\uvect{n}\) is not constant around that ring.  The
virtual-emission depth is therefore not uniform along the ring.  Points
on the side of the ring toward burst A have larger
\(\uvect{b}_B\cdot\uvect{n}\), and hence correspond to larger depths,
whereas points on the opposite side are shallower.  The depth
distribution for any chosen pair and observing time can be computed
directly from Eq.~\eqref{eq:s_of_n}.

For pair selection, it is useful to introduce a characteristic depth
scale by evaluating the denominator at the burst separation angle:
\begin{equation}
    s_B
    \equiv
    {c\Delta\over 1-\cos\alpha}
    \simeq
    {2c\Delta\over \alpha^2}.
    \label{eq:sB_def}
\end{equation}
Numerically,
\begin{equation}
    s_B
    \simeq
    20.1\,{\rm kpc}
    \left({\Delta\over 10\,{\rm yr}}\right)
    \left({\alpha\over 1^\circ}\right)^{-2}.
    \label{eq:sB_numeric}
\end{equation}
Thus decade-separated burst pairs with angular separations of order
one degree naturally probe Galactic-scale distances.

The scale in Eq.~\eqref{eq:sB_numeric} is not merely formal.  Existing
GRB records include pairs satisfying these criteria.  As an illustrative
check, Table~\ref{tab:alpha_one_pairs} lists examples with
\(0.9^\circ\le\alpha\le1.1^\circ\),
\(\Delta\ge10\,{\rm yr}\), and
\(\sigma_A,\sigma_B\le1''\), where \(\sigma_A\) and \(\sigma_B\) denote
the adopted angular localization uncertainties of the two bursts.  The
examples are selected from a working catalog constructed from the public
GRBWeb database.

\begin{table}
\centering
\caption{
Illustrative GRB pairs near \(\alpha\simeq1^\circ\).
The pairs satisfy
\(0.9^\circ\le\alpha\le1.1^\circ\),
\(\Delta\ge10\,{\rm yr}\), and
\(\sigma_A,\sigma_B\le1''\) in the working catalog used here.  The
column \(b_B\) gives the Galactic latitude of the later burst, which
approximates the early sky region sampled by the ring.
}
\label{tab:alpha_one_pairs}
\setlength{\tabcolsep}{4pt}
\begin{tabular}{lccccc}
\hline
Pair
& \(\Delta t\)
& \(\alpha\)
& \(s_B\)
& \(b_B\)
& \(\sigma_A,\sigma_B\) \\
&
\(({\rm yr})\)
& \(({\rm deg})\)
& \(({\rm kpc})\)
& \(({\rm deg})\)
& \(({\rm arcsec})\) \\
\hline
061222B--251103B & 18.87 & 1.094 & 31.7 & \(-8.6\)  & 0.79,\ 0.93 \\
081102A--230409B & 14.43 & 1.013 & 28.3 & \(-1.6\)  & 0.70,\ 0.93 \\
060512A--190211A & 12.75 & 1.021 & 24.6 & \(+74.8\) & 0.23,\ 0.25 \\
081121A--200630A & 11.60 & 1.017 & 22.6 & \(-29.0\) & 0.28,\ 0.93 \\
090726A--200829A & 11.09 & 0.937 & 25.5 & \(+35.2\) & 0.70,\ 0.16 \\
\hline
\end{tabular}
\end{table}

\section{Error budget}
\label{sec:error}

For an observational use of the prescription, one chooses an ordered
burst pair \((A,B)\) with \(t_B>t_A\), adopts the observed burst
directions \(\uvect{b}_A\), \(\uvect{b}_B\), and computes the
receiver-frame arrival-time separation \(\Delta=t_B-t_A\).  For a chosen
observing epoch \(T=t_B+\tau\), the delay \(\tau\) fixes the
dimensionless time \(u=\tau/\Delta\).  Substituting these quantities
into Eq.~\eqref{eq:ring_exact} gives the corresponding angular search
ring.  Timing errors in the burst arrival times affect the ring only
through \(u\), and are typically much smaller than the
angular-localization errors considered here.  The dominant geometrical
uncertainty is therefore expected to come from the angular localizations
of the two bursts.

The small-angle form of Section~\ref{sec:ring_depth} makes the error
scaling transparent.  The burst-position errors affect the search ring
in two geometrically distinct ways: they shift the ring center on the
sky and they change the ring radius.  Keeping only the leading terms in
the tangent-plane approximation, the center uncertainty is
\begin{equation}
    \sigma_c^2(u)
    \simeq
    (1+u)^2\sigma_B^2+u^2\sigma_A^2 .
    \label{eq:center_error}
\end{equation}
The corresponding radial uncertainty is
\begin{equation}
    \sigma_\rho^2(u)
    \simeq
    u(1+u)\sigma_\alpha^2 ,
    \qquad
    \sigma_\alpha^2\simeq \sigma_A^2+\sigma_B^2 ,
    \label{eq:radius_error}
\end{equation}
where \(\sigma_A\) and \(\sigma_B\) are the characteristic angular
localization errors of the two bursts, and \(\sigma_\alpha\) denotes the
uncertainty in their angular separation, approximated here by adding the
independent projected errors in quadrature.  Thus, for comparable burst
localizations, the characteristic width of the search ring is of order
\((1+u)\sigma_\alpha\), up to factors of order unity.  The key point is
that the width is controlled by burst localization, not by any
astrophysical distance uncertainty.

It is useful to compare this scale with the finite-anchor hybrid
construction of \citet{Seto2026}.  In the GRB--TESS application
discussed there, the search ring has a radius of order
\(\rho\simeq0.8^\circ\), and its width is mainly controlled by the
fractional uncertainty in the Galactic-center distance.  With
\(\delta D_{\rm GC}/D_{\rm GC}\simeq5\times10^{-3}\)
\citep{2021A&A...647A..59G}, this gives an anchor-induced angular width
of order
\begin{equation}
    {1\over2}\rho\,{\delta D_{\rm GC}\over D_{\rm GC}}
    \sim 10'' .
\end{equation}
By contrast, in the present dual-burst construction, the corresponding
width is set by the burst localizations.  For well-localized pairs with
subarcsecond burst positions and \(u=O(1)\), the characteristic width can
therefore be substantially narrower.

The plane-wave approximation introduces one additional systematic check.
For a burst at finite distance \(R\), the incoming wavefront is
spherical rather than exactly planar.  Expanding the corrected
arrival-time function for \(s\ll R\), where \(s\) is the
virtual-emission distance, gives a characteristic angular displacement
of order
\begin{equation}
    \delta\theta_{\rm curv}
    \sim
    {s\theta\over2R}
    \simeq
    0.36\,{\rm arcsec}
    \left({s\over20\,{\rm kpc}}\right)
    \left({\theta\over1^\circ}\right)
    \left({R\over100\,{\rm Mpc}}\right)^{-1},
    \label{eq:curvature_error}
\end{equation}
where \(\theta\) is the angular offset from the burst direction.  For
typical cosmological GRBs, \(R\) is much larger than
\(100\,{\rm Mpc}\), so the plane-wave approximation has a large margin.
Even at \(R\sim100\,{\rm Mpc}\), the curvature-induced displacement in
Eq.~\eqref{eq:curvature_error} remains below the arcsecond scale for the
Galactic-scale depths and degree-scale angular offsets considered here.
When the distance to a reference burst is known, the same calculation
can be repeated with the corresponding spherical wavefront if a narrower
search mask requires it.  This provides a controlled finite-distance
extension of the plane-wave approximation.

\section{Search strategies}
\label{sec:strategies}

The dual-burst prescription removes the need for a finite-distance
spatial anchor, but it leaves a new Schelling-like problem: the selection
of the burst pair itself.  If a catalog contains \(N\) usable bursts,
there are \(N(N-1)/2\) possible pairs, so an arbitrary pair choice is not
a plausible basis for implicit coordination.  A useful strategy must
therefore reduce this pair space by a rule that is simple,
reproducible, and based on information available in common burst
records.

We focus on two such reductions: anchoring the pair to an exceptionally
salient burst, and selecting pairs whose geometry is favorable for
Galactic searches.

\subsection{Salient-burst anchored pairs}
\label{subsec:boat_pairs}

The most direct way to reduce the burst-pair selection problem is to
require one or both members of the pair to be exceptionally salient.
This converts an unrestricted choice among many possible pairs into a
smaller set anchored by conspicuous events.  Such anchors are attractive
as Schelling references because they can be selected independently from
commonly available burst records.

GRB~221009A provides the clearest current example.  Its observed fluence
was far above that of other well-documented GRBs
\citep{2023ApJ...946L..31B,2023ApJ...949L...7F,2023ApJ...952L..42L},
motivating its description as the ``brightest of all time'' (BOAT) GRB
\citep{2023ApJ...946L..31B}.  This exceptional brightness corresponds
to an inferred rarity of order one event per \(10^4\) yr, making it a
particularly natural Schelling anchor
\citep{Seto:2025iul}.  Its low Galactic latitude,
\(b\simeq4.3^\circ\), further makes it attractive for searches involving
dense Galactic stellar fields.

\subsection{Geometrically favorable pair selection}
\label{subsec:geometrical_pairs}

The geometrical role of pair selection is to tune the search ring to the
Galactic stellar distribution.  The key parameters are the angular
separation \(\alpha\) and the time separation \(\Delta\).  As shown in
Section~\ref{sec:ring_depth}, these control the characteristic
virtual-emission depth,
\begin{equation}
    s_B
    \simeq
    {2c\Delta\over\alpha^2}
    \simeq
    20.1\,{\rm kpc}
    \left({\Delta\over10\,{\rm yr}}\right)
    \left({\alpha\over1^\circ}\right)^{-2},
    \label{eq:sB_strategy}
\end{equation}
and, at fixed \(u=\tau/\Delta\), the angular size of the ring,
\begin{equation}
    \rho_{\rm ring}
    =
    \alpha
    \sqrt{
    {\tau\over\Delta}
    \left(1+{\tau\over\Delta}\right)
    } .
    \label{eq:ring_growth_strategy}
\end{equation}
Thus smaller angular separations give deeper and narrower rings, whereas
larger angular separations give shallower and wider rings.  The time
separation \(\Delta\) sets the overall depth scale and determines how a
given observing delay maps to \(u\).

The favorable choice of \(\alpha\) and \(\Delta\) is therefore not
universal.  Toward the Galactic disk, increasing the characteristic
depth can sample dense stellar regions over a longer path length.  At
high Galactic latitude, by contrast, the line of sight quickly leaves the
thin disk, so increasing \(s_B\) does not necessarily add many Galactic
targets.  In such directions, angular coverage and ring motion can be
more important than depth alone.  A practical pair
ranking should therefore combine the geometric scalings above with the
Galactic stellar distribution and the available time-domain survey
coverage.

A complementary use of the same pair-selection framework is to start
from a specific line of sight rather than from a sky ring.  For example,
one may ask whether a ring generated by a focal burst pair crosses the
direction of a known astrobiologically interesting stellar target, and
at what observing epoch this occurs.  This target-specific version still
uses only the stellar direction, not the stellar distance, and therefore
retains the distance-free character of the prescription.  Such
applications form a natural search-design extension of the present
geometrical prescription, connecting it to targeted optical or radio SETI
searches where particular stellar directions and observing windows must
be prioritized.

\section{Summary and discussion}
\label{sec:summary}

We proposed a dual-burst SETI prescription in which two distant
astrophysical transients provide a distance-free reference structure for
concurrent signaling.  The local encounters of the two burst wavefronts
define reference events, and outgoing virtual rays from those events
define the corresponding signal trajectories.  In the plane-wave
approximation, the resulting search ring, Eq.~\eqref{eq:ring_exact},
depends only on the two burst directions, their observed arrival-time
separation, and the chosen observing time, not on the distances to the
bursts, to a possible transmitter, or to any Galactic spatial anchor.

We derived the ring geometry, the associated virtual-emission depth, and
the leading error budget.  The characteristic depth scale
\(s_B\simeq2c\Delta/\alpha^2\) shows that decade-separated GRB pairs
with degree-scale angular separations naturally probe Galactic-scale
distances.  For suitably well-localized pairs, the leading ring width is set by
burst localization rather than by any astrophysical distance
uncertainty, allowing arcsecond-scale search rings in favorable cases.

The removal of a finite-distance spatial anchor shifts the remaining
Schelling problem to burst-pair selection.  We emphasized that the pair
should be chosen using simple and reproducible criteria, such as the
salience of an exceptionally bright burst, the angular and temporal
separations of the pair, localization quality, and the Galactic stellar
distribution sampled by the resulting ring.  GRB~221009A provides a
natural example of a salient anchor, while existing GRB records include decade-separated pairs satisfying
the illustrative Galactic-depth criteria discussed above.

The formulae above are intended as the leading geometrical prescription.
When constructing search masks substantially narrower than an arcsecond,
one can use the usual high-precision astrometric and time-standard
procedures, including relativistic corrections where needed.

\begin{acknowledgments}
The author thanks Y. Uno for useful conversations.
\end{acknowledgments}
\bibliographystyle{aasjournal}
\bibliography{dual_burst}

\end{document}